\documentclass[lettersize,journal]{IEEEtran}
\usepackage{amsmath,amsfonts}
\usepackage{algorithmic}
\usepackage{algorithm}
\usepackage{array}
\usepackage{textcomp}
\usepackage{stfloats}
\usepackage{url}
\usepackage{verbatim}
\usepackage{graphicx}
\usepackage{cite}
\usepackage{color}
\usepackage{amssymb}
\usepackage{tabularx}
\usepackage{pifont}
\usepackage{lipsum}
\usepackage{caption}

\usepackage{subfigure}
\usepackage{booktabs} 

\begin{document}

\title{Digital Twin-Empowered Task Assignment in Aerial MEC Network: A Resource Coalition Cooperation Approach with Generative Model}

\author{Xin Tang, Qian Chen, Rong Yu, Xiaohuan Li
\thanks{This work was supported in part by the Science and Technology Key Project of Guangxi Province (AB23026038), in part by the National Natural Science Foundation of China under Grants (U22A2054), in part by the Graduate Study Abroad Program of Guilin University of Electronic Technology (GDYX2024001). {\itshape (Corresponding author: Xiaohuan Li).}}
\thanks{X. Tang, X. Li are with the Guangxi University Key Laboratory of Intelligent Networking and Scenario System (School of Information and Communication, Guilin University of Electronic Technology), Guilin, 541004, China, and also with National Engineering Laboratory for Comprehensive Transportation Big Data Application Technology (Guangxi), Nanning, 530001, China (e-mails: tangx@mails.guet.edu.cn; lxhguet@guet.edu.cn).}
\thanks{Q. Chen is with the School of Architecture and Transportation Engineering, Guilin University of Electronic Technology, Guilin, 541004, China (e-mails: chenqian@mails.guet.edu.cn).}
\thanks{R. Yu is with School of Automation, Guangdong University of Technology, Guangzhou, 510006, China (e-mail: yurong@ieee.org).}
}
\markboth{IEEE Transactions on Network Science and Engineering, Citation information: DOI 10.1109/TNSE.2024.3482327}%
{Shell \MakeLowercase{\textit{et al.}}: A Sample Paper Using IEEE tran.cls for IEEE Journals}

\maketitle

\begin{abstract}
To meet the demands for ubiquitous communication and temporary edge computing in 6G networks, aerial mobile edge computing (MEC) networks have been envisioned as a new paradigm. However, dynamic user requests pose challenges for task assignment strategies. Most of the existing research assumes that the strategy is deployed on ground base station (GBS) or unmanned aerial vehicle (UAV), which will be ineffective in an environment lacking infrastructure and continuous energy supply. Moreover, the resource mutual exclusion problem of dynamic task assignment has not been effectively solved. Toward this end, we introduce the digital twin (DT) into the aerial MEC network to study the resource coalition cooperation approach with the generative model (GM), which provides a preliminary coalition structure for the coalition game. Specifically, we propose a novel network framework that is composed of an application plane, a physical plane, and a virtual plane. After that, the task assignment problem is simplified to convex optimization programming with linear constraints. And then, we also propose a resource coalition cooperation approach that is based on a transferable utility (TU) coalition game to obtain an approximate optimal solution. Extensive analysis and numerical results confirm the effectiveness of our proposed approach in terms of energy consumption and utilization of resources.
\end{abstract}

\begin{IEEEkeywords}
resource cooperation, task assignment, digital twin (DT), coalition game, generative model (GM), unmanned aerial vehicle (UAV), mobile edge computing (MEC)
\end{IEEEkeywords}

\section{Introduction}
\begin{figure}[!t]
	\centering
	\includegraphics[width=3.5in]{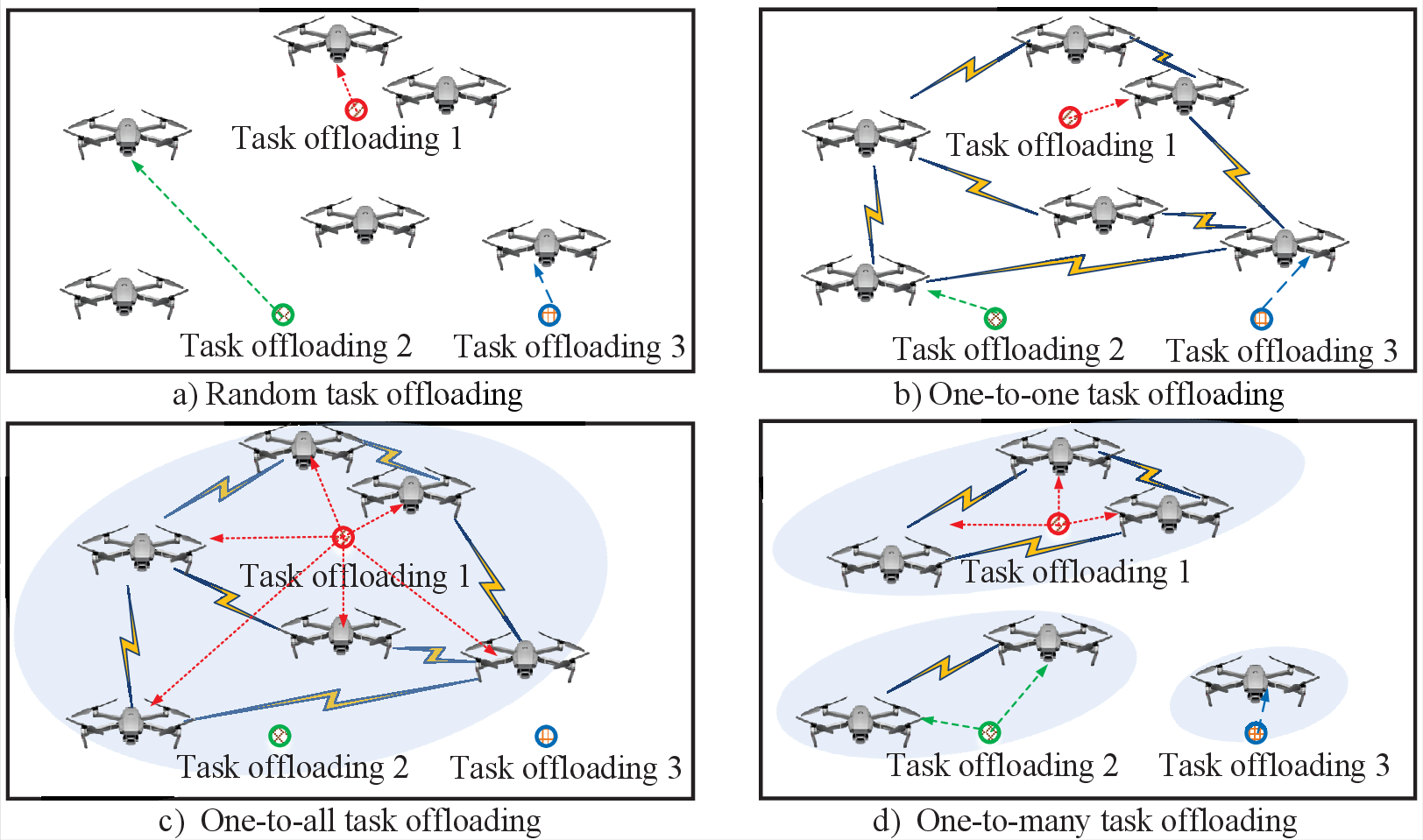}%
	\caption{Four task offloading models in a multi-UAV network with high-dynamic task requests and uneven onboard resource distribution.}
	\label{fig01}
\end{figure}

The increasing demand for surveillance and remote sensing in disasters or remote environments, such as maritime areas, deserts, and dense forests, necessitates revolutionary solutions. Implementing mobile services and running intelligent applications on mobile edge devices (MEDs), like mobile phones, unmanned surface vehicles (USVs), and unmanned ground vehicles (UGVs), offers such a revolutionary approach \cite{9372288, 9865231}. Nevertheless, the computing and communication capacity of a MED is still not comparable with that of the CPU or GPU mounted on the static ground base station (GBS) \cite{huang2019cloud, sun2019adaptive}. Admittedly, there are still some deficiencies to be tackled, i.e., the location limitations of the GBS and the high deployment cost. Owing to the flexible deployment and low price of aerial crafts, such as unmanned aerial vehicles (UAVs), airships or balloons \cite{jing2023cooperative, chen2023enhancing}, the aircraft as mobile edge computing (MEC) devices has emerged as a key role for providing computing power services to users.

Several studies have been dedicated to aircraft-assisted MED for aerial mobile edge services, such as task offloading \cite{cheng2023ai, chen2021multi}. The researchers mainly focus on reducing computing delay and energy consumption. However, since computing, cache, bandwidth, and other resources of the aircraft serving dynamic task offloading requests are unevenly distributed among multiple aircrafts, mutual exclusion problems in limited resource can easily arise. The existing research primarily solves the problems by using artificial intelligence (AI) and game theory. The AI algorithms train deep neural networks to provide near-optimal policies for task offloading, while the intelligent algorithms lack interpretability and generalization. The game theory converts the interaction of multiple aircrafts into a game, and the mutual exclusion problem is resolved by utilizing equilibrium conditions of the game, i.e., encouraging resource cooperation among aircrafts.

\begin{table*}[!t]
  \centering
  \caption{Comparison between related works and our work \label{tab:table00}} 
  \begin{tabular*}{\textwidth}{@{\extracolsep{\fill}}|c|c|c|c|c|c|c|} 
    \hline
      & \multicolumn{3}{c|}{Aerial MEC network architecture} & \multicolumn{3}{c|}{Optimization objective} \\
    \hline
    Ref. &One-to-many offloading &HAP-assistance &UAVs cooperation &U2U interaction &Energy consumption &Resource utilization \\
    \hline
   \cite{sun2024multi} &{}&{}&{ }&{ }& {\checkmark} & \\
    \hline
    \cite{liu2024latency} &{\checkmark}&{ }&{ }&{ }& {\checkmark} & \\
    \hline
    \cite{chen2020joint} &{ }&{ }& {\checkmark} & {\checkmark} & & {\checkmark} \\
    \hline
    \cite{lyu2023computing} &{ }&{\checkmark}&{ }&{ }&{ }& {\checkmark} \\
    \hline
    \cite{ju2024multi} &{\checkmark}& & {\checkmark} & {\checkmark} & {\checkmark} &{ }\\
    \hline
    \cite{bai2022delay} & & & {\checkmark} & {\checkmark} & & {\checkmark} \\
    \hline
    \cite{qi2022task} &{ }&{ }& {\checkmark} & {\checkmark} & {\checkmark} & \\
    \hline
    \cite{liu2024edge} &{ }& {\checkmark} & {\checkmark} & {\checkmark} & {\checkmark} &{ }\\
    \hline
    \cite{sun2024joint} &{ }& {\checkmark} & & & & \\
    \hline
    \cite{alam2024joint} &{ }& {\checkmark} & & & {\checkmark} &{ }\\
    \hline
    Our work & {\checkmark} & {\checkmark} & {\checkmark} & {\checkmark} & {\checkmark} & {\checkmark} \\
    \hline
  \end{tabular*}
\end{table*}

In order to better introduce the motivation of this paper, we summarize existing task offloading methods into the following four categories. The examples in Fig. \ref{fig01} are given below to illustrate the challenges that existing multi-UAV systems deal with task offloading. In Fig. \ref{fig01} a), tasks are offloaded to UAVs via MED to UAV (M2U) communications randomly or selectively without considering the multi-UAV cooperation. This often results in poor task offloading service quality and inefficient resource utilization due to resource shortages or surpluses. Related to this task offloading schedule under the multi-UAV system, in \cite{sun2024multi}, the authors consider a multi-UAV-assisted MEC system. Each user could generate a task that could be offloaded to a UAV. And the authors propose a joint task offloading, computation resource allocation, and UAV trajectory control approach to solve a multi-objective optimization problem. In \cite{liu2024latency}, the authors design the task for each user, which can be divided into two subtasks. Each user can execute one of its subtasks locally and offload the other to a UAV or GBS. Fig. \ref{fig01} b) illustrates the use of the equilibrium method among multi-UAV providing one-to-one offloading services. Each UAV provides task offloading services independently, requiring careful evaluation of the inter-UAV relationships by UAV to UAV (U2U) communications. However, this method leads to unjust resource distribution and, subsequently, lower task offloading service quality. In \cite{chen2020joint}, the authors propose a cooperative reconnaissance task and spectrum access scheme for coalition-based UAV networks and prove that the proposed scheme can converge to a stable coalition while maximizing network utility. In \cite{lyu2023computing}, the authors propose UAV-assisted emergency communication with nonorthogonal multiple access and present a joint offloading decision and resource allocation problem to minimize computation overhead. Fig. \ref{fig01} c) shows a scenario where all UAVs cooperate with each other to form a multi-UAV system to provide all-to-one task offloading services. The approach enhances the quality of task offloading services, it can also lead to underutilization of UAV resources. The representative scenario of \cite{9390406} is similar to this case, all the cellular users cooperate with each other for mobile data forwarding task. In \cite{ju2024multi}, the authors design a multi-UAV-assisted MEC system with software-defined networking technology to provide task offloading services for edge user devices within the region. The aim is to optimize UAV flight trajectories and minimize overall system energy consumption. In \cite{bai2022delay}, the authors propose two possible sources of the tasks in practical scenarios, i.e., ground users and UAVs, and investigate the delay-aware cooperative task offloading problem for the multi-UAV-enabled edge cloud computing system. In contrast, Fig. \ref{fig01} d) differs from Fig. \ref{fig01} c) by simultaneously considering task offloading service requests and resource cooperation. Multiple UAVs to form a team with the aim of providing many-to-one task offloading services while maximizing overall utility. This method excels in optimizing both resource utilization and task offloading service quality. We clearly see that the method described in Fig. \ref{fig01} d) is more in line with actual needs. In \cite{qi2022task}, the authors investigate overlapping coalition formation in a task-driven cooperative UAV network. The resources of UAVs are allocated to the same task to form a coalition to execute the task cooperatively. Numerical results show that the proposed algorithm increases the average utility of tasks. In \cite{liu2024edge}, the authors propose an edge computing network architecture for space-air-ground integration and utilize a binary integer linear programming model to resolve the offloading cost minimization problem. Furthermore, they prove the existence of Nash equilibrium through the potential game. However, the UAV-assisted MED also faces new challenges, i.e., the long line-of-sight link between the user and the UAV will cause a long transmission latency, and the computing resources at the UAV are not always adequate, which are not friendly to delay-sensitive tasks \cite{liu2021joint}. Although some excellent work has demonstrated that device-to-device task offloading can be used to ensure low latency and high data transmission rate \cite{lai2022adaptive, hoa2023dynamic, ozer2023offloading}, the service requirements of user and the unpredictability of UAV resources in the real world have not yet been fully considered. And it can also be seen from the above four examples that the task offloading in an aerial MEC network is typically a dynamic process. The real-time task offloading will pose a great challenge to the dynamic resource. So it is mutual exclusion with the limited resources of the aerial MEC system. Therefore, the resource cooperation approach in the aerial MEC network to execute task offloading based on a coalition formation rule needs to be further investigated. To clearly illustrate the difference between the resource coalition cooperation approach and the related works in this field, a comparison in terms of aerial MEC network architecture and optimization objective are provided in Table \ref{tab:table00}. If the performance metric satisfies the corresponding attribute, it is marked with $\checkmark$.

The emergence of digital twin (DT) technology enabled the cooperation and information exchange between the physical object and virtual entity \cite{tao2018digital}. DT not only generates virtual models representing physical objects within the network but also maintains real-time monitoring of the network's status. This enables the direct provisioning of facilitating more accurate and timely offloading decisions to users, aligning with the evolving needs of intelligent systems \cite{guo2022federated}. Recognizing the advantages of DT, recent researches have combined DT and MEC to establish an aerial MEC network that collects data from diverse physical entities and stores the data in a dedicated device while simultaneously conducting monitoring of the current network's status. On the other hand, the generative model (GM) provides an exciting perspective for intention-based resource cooperation strategy, where we can encode the intention into the conditioning information and explicitly guide the generation of cooperation strategy (i.e., request-strategy-utility sequences that represent the decision-making process) with flexible combination of multiple task constraints.

With respect to the system model, most of the aforementioned use one or more UAVs to provide resources in aerial MEC networks \cite{zhang2022deep, luo2023decentralized, li2023robust}. However, a UAV's resources are limited, making it challenging to support the server requirements of large-scale tasks from MEDs. Although GBS can provide auxiliary resources, they are often unavailable in emergency scenarios such as military attacks or natural disasters. To address this challenge, some researchers concentrate on the integration of high-altitude platform (HAP) into aerial MEC systems \cite{sun2024joint, alam2024joint}. By establishing effective cooperation between UAVs and HAP, less additional U2U interaction is required when generating resource cooperation strategies. Inspired by the motivation and the challenge, the emerging technique and practical scenarios, this paper considers DT-empowered task assignment with resource coalition cooperation in an aerial MEC network consisting of multiple users, UAVs equipped with edge servers, and an airship equipped with a more powerful cloud server, i.e., HAP. In this case, our goal is to reduce energy consumption by jointly optimizing the task assignment, energy consumption, communication bandwidth, and the resource utilization. For the sake of clarity, the main contributions of this work are listed as follows
\begin{itemize}
\item{We propose a DT-empowered aerial MEC network for task assignment based on resource coalition cooperation, which comprises a physical plane, a virtual plane, and an application plane. The virtual plane is responsible for collecting status information from devices in the physical plane to build a resource pool through UAV-to-airship (U2A) communications, and subsequently generating the resource cooperation strategy for task assignment.}

\item{We study the joint optimization problem of resource utilization, coalition utility, and energy consumption, considering multiple constraints, i.e., the assigning capacity of tasks, latency constraints, bandwidth, and cache. And then we transform this non-convex resource optimization problem into a convex optimization problem with linear constraints.}

\item{We present a resource coalition cooperation approach that is based on a transferable utility (TU) coalition game. The approach uses an iterative solution to optimize the utilization of resources in the UAV coalition. Furthermore, the GM can generate a preliminary coalition structure, which can be directly applied to the coalition game to reduce iterations and further reduce energy consumption.}
\end{itemize}

The rest of the section-wise paper outline is detailed as follows. Section \ref{sec:2} presents a brief synopsis of recent literature in this area. Section \ref{sec:3} introduces the system architecture and model. The question statement is presented in Section \ref{sec:4}. A novel resource coalition cooperation approach is given in Section \ref{sec:5}. Section \ref{sec:6} analyzes the performance of the proposed approach. The simulation results are discussed in Section \ref{sec:7} and conclude this paper in Section \ref{sec:8}.

\section{Related Work}
\label{sec:2}
In this section, we review research on the integration of aircraft, DT, and GM to meet the new demands of future networks, such as ubiquitous edge computing, intelligent resource allocation, robust task assignment, and network design.

\subsection{Aerial-Assisted MEC Networks}
Recently, aerial with high mobility has received increased attention in various application scenarios, such as MEC networks and IoT. Focusing on optimizing UAV trajectory, resource allocation, and task scheduling, the authors in \cite{zhang2022deep} propose two deep reinforcement learning (DRL) algorithms to maximize the average aggregate quality of experience of all IoT devices with UAV assistance by considering user assignment, UAV trajectory, bandwidth, and computing resource allocation. The authors in \cite{luo2023decentralized} use UAV to assist ground users to offload the task and process the data on edge in which a decentralized user allocation and dynamic service scheme are proposed for assigning the UAV deployment, and the training of the DRL is divided into lower layer and upper layer. The authors in \cite{li2023robust} formulate a multi-agent Markov decision process and propose a multi-agent proximal policy optimization with Beta distribution framework to solve with the coupled optimization variables and the high uncertainties in the MEC network. However, these studies focus on a single UAV scenario means that there only exists one server or relay. When the number of tasks increases, resource shortages and bandwidth constraints will occur.

Extensive research has shown that multi-UAV can also play an important role in MEC networks. A series of studies have explored how to improve the performance of the MEC networks in terms of energy consumption, latency, and total cost. The authors in \cite{hoa2023deep} propose an UAV-assisted MEC system in which a user can offload its task to multi-UAV in an ad-hoc fashion for the task execution can be finished faster. The authors in \cite{zhou2023priority} investigate the joint impact of task priority and mobile computing service on the MEC networks. They employ a DRL algorithm to learn an effective solution by continuous interaction between the UAV agent and environment. Fortunately, game theory provides the ideal framework for designing efficient and robust distributed algorithms. More recently, there has been growing interest in integrating game theory with MEC networks. In \cite{sun2021dynamic}, the authors establish a dynamic DT of UAV-assisted Internet of vehicle and design a two-stage incentive mechanism for resource allocation based on the Stackelberg game to capture the time-varying resource supply and demands so that resource scheduling can be performed. However, these studies require a large number of interactions between nodes to obtain network information, which leads to increased task execution delays and energy consumption. Moreover, the resource cooperation method between nodes is not considered, and the resource utilization efficiency is low.

\subsection{DT-empowered Resource Allocation and Task Assignment}
The issue of resource limitation remains a critical concern within aerial MEC networks. To enhance resource utilization, DT technology has emerged as a pivotal solution for managing resources in these networks. DT has been widely adopted in optimizing resource management for efficient MEC. In \cite{liu2023energy}, the authors propose a MEC network with multi-UAV and a DT-empowered GBS to enhance the performance of the MEC for mobile users. The authors in \cite{guo2023resource} propose a DT-empowered dynamic resource allocation strategy to meet various resource allocation requirements for users in a UAV-assisted mobile network. In \cite{li2023flexedge}, the authors study DT-enabled task offloading method for the UAV-aided vehicular edge computing networks to achieve computing resource management. These studies propose useful insights into the air-ground network, but they mainly focus on the ground DT.

More recently, there has been growing interest in the user mobility and unpredictable MEC environments. The authors in \cite{guo2023intelligent} propose DT-empowered UAV deployment and hybrid task offloading strategies to achieve global optimal resource allocation in aerial computing networks. Similarly, the authors in \cite{li2022digital} study the intelligent task offloading method in an UAV-enabled MEC network with the assistance of DT. They aim to minimize the energy consumption of the entire MEC network by jointly optimizing UAV trajectory, transmission power distribution, and computation capacity allocation while considering the constraints of task maximum processing delays. In \cite{wang2023digital}, the researchers exploit the real-time prediction capability of DT to optimize the computation offloading decision and meet the high delay sensitivity requirements under the mismatch between the sudden demands of users and the limited computation resources of UAVs. These works have proved that the research on DT and multi-UAV is valuable and significant in terms of improving the performance of MEC systems. However, while these works use DRL, matching theory, and DT technology to achieve optimal resource allocation and task offloading strategies, they do not consider establishing a unified resource pool or utility-maximizing resource providers.
\begin{figure*}[htb]
    \centering
	\includegraphics[width=\textwidth]{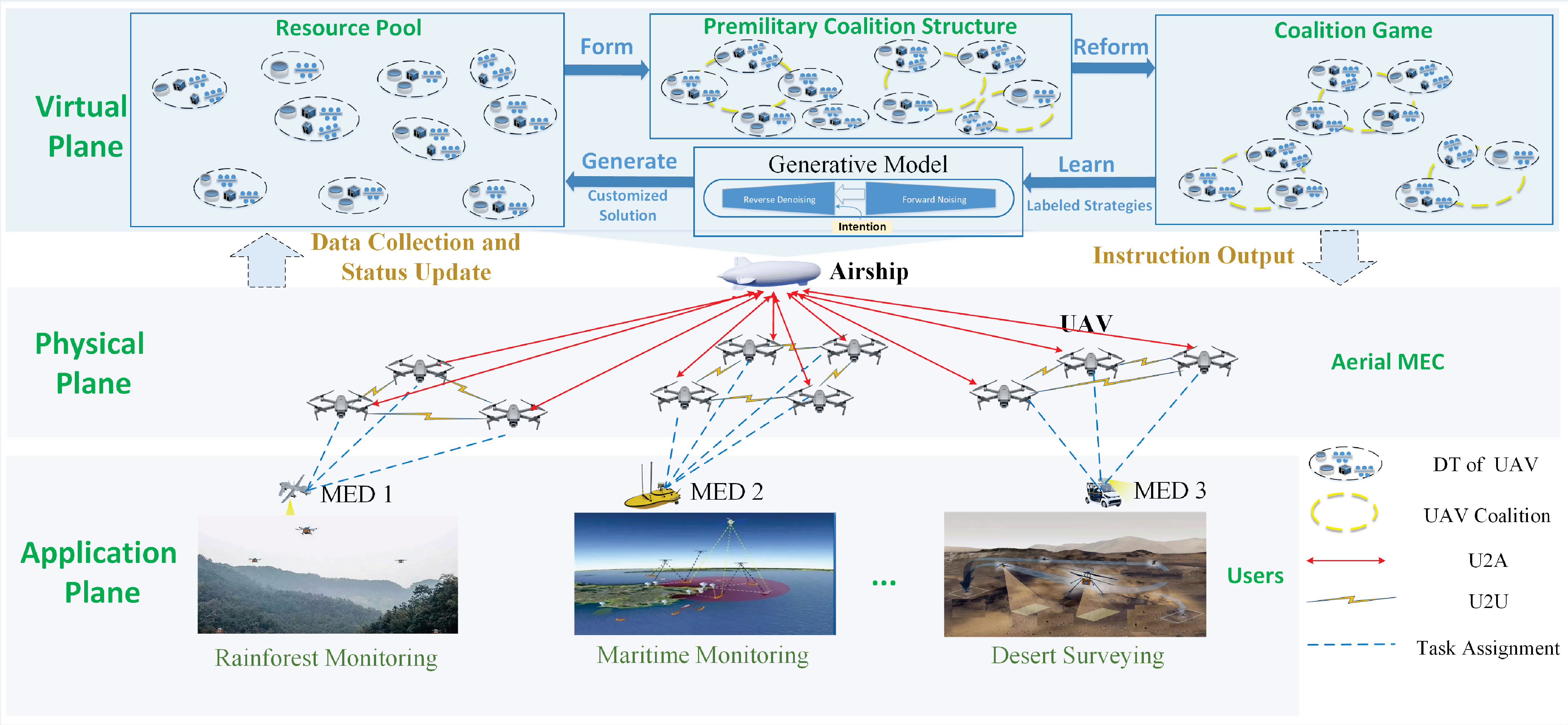}%
	\caption{System model.}
	\label{fig02}
\end{figure*}
\subsection{Application of GM in Edge Intelligence Network}
GMs have emerged as a promising technology to improve the efficiency, diversity and flexibility of the content generation process by adopting a variety of AI models. Deploying generative artificial intelligent (GAI) models in intelligence networks has been expected to enhance the quality of service. The authors in \cite{liu2023optimizing} present the concept of mobile-edge AI-generated everything and discussed its optimization using prompt engineering and illustrate how much improvement prompt engineering can lead to in terms of quality of experiment, service latency, and bandwidth usage by a case study. Furthermore, diffusion models showed great potential in optimization since the denoising process starting from Gaussian noise can be regarded as the optimization process of finding the optimal solution. In \cite{huang2023ai}, the authors propose a novel intention-driven GAI paradigm for network design, which can quickly generate a variety of customized solutions. The proposed method treats labeled trajectories in online reinforcement learning (RL) as the original data distributions, and then the offline RL with the diffusion model has the ability to operate in high-dimensional state spaces and learn new customized strategies that closely match the expected network design. The authors in \cite{du2023beyond} introduce a diffusion model to deep DRL, which greatly improves the exploration ability and achieves the better performance in many optimization problems, such as incentive mechanism design, resource allocation, and channel coding for wireless networks.

To the best of our knowledge, this is the first time we have researched DT-empowered task assignment by coalition game with a GM in an aerial MEC network. Specifically, the system model can monitor the status of the network in real-time and use the method of resource coalition cooperation to optimize task assignment and execution efficiency.

\section{System Architecture and Model}
In this section, we first propose the aerial MEC system architecture for task assignment. Then, we formulate the DT model, communication model and energy consumption model.

\subsection{System Architecture}
\label{sec:3}
This system is designed to use a HAP (e.g., airship) with high computing power, large caches, and strong communications to assist multi-UAV in completing resource optimization to adapt to MED's task assignment requests. We build DT models of UAVs in an airship to optimize the resource cooperation of aerial MEC network. With the help of DT, the system model designed in this paper is shown in Fig. \ref{fig02}, which is composed of an application plane, a physical plane, and a virtual plane. The model parameters are shown in Table \ref{tab:table1}. Next, we describe the system model as follows.

\textbf{Application plane:} This plane involves users (e.g., road inspection vehicles, reconnaissance aircraft, and smart ships) sending task assignment requests and executing the task assignment. Specifically, the tasks are offloaded to specific UAV coalitions based on resource cooperation strategies.

\textbf{Physical plane:} This plane encompasses three main functions. First, it relays the UAVs' status information and users' task assignment requests to the airship. Second, the UAVs receive a resource collaboration strategy from the airship, enabling them to form the necessary coalitions. Third, the UAVs establish an edge computing network to provide computing services to users.

\textbf{Virtual plane:} After the airship receives the intentions (e.g., the task assignment requests), it generates samples from high-utility strategies and infers a customized cooperative solution for the resource pool. And then the preliminary coalition structure is formed. Furthermore, the preliminary coalition structure could be directly applied to the coalition game to quickly work out the optional coalition cooperation strategies. The strategies are viable to make into labeled strategy datasets for particular task assignments and are stored in a standardized manner for model training. In addition, the historical data of other airship systems can be encoded into labeled strategies to further improve the quality of the datasets. Note that the GM is not limited to the diffusion model, and other models could also be applied to this proposed approach.

\subsection{DT Model}
In this paper, the M2U is also designed to obtain the MED task parameters to construct a DT model. In addition, the communication between UAV and airship (U2A) is designed to transmit the DT modeling parameters of MED and UAV. The above communication process all adopts the communication method of time division multiple access. To simplify the analysis, this paper defines the system model as a synchronous periodic time slot system, and the duration of each time slot $t$ is $\tau$. Let the MED set be $M = \left\{{1, 2,\cdots, i, \cdots, m} \right\}$. $m$ means the maximum sequence number of MED. The $i$th MED is represented as $MED_{i}$. Then the DT of $i$th MED in time slot $t$ is
\begin{equation}
	DT^{t}_{MED_{i}} = \{l^{t}_{MED_{i}},s^{t}_{i},\vartheta^{t}_{i}, p^{t}_{i,tr}, \tau^{t}_{i} \},
\end{equation}
where $l^{t}_{MED_{i}}$ represents coordinate $(x^{t}_{i},y^{t}_{i},z^{t}_{i})$ of $MED_{i}$. $s^{t}_{i}$ is the amount of tasks that $MED_{i}$ requests to assign within the time slot $t$. $\vartheta^{t}_{i}$ represents the computing resource required to process the amount of tasks per unit, that is, computational complexity. $p^{t}_{i,tr}$ represents the transmission power of $MED_{i}$ for task assignment. $\tau^{t}_{i}$ means the time slot constraint of task, i.e., the maximum time allowed to complete the task of $MED_{i}$, which is $\tau^{t}_{i}\leq\tau$.

The set of UAV is denoted as $V = \left\{ {1, 2, \cdots, j, \cdots, n} \right\}$. $n$ means the maximum sequence number of UAV. Let $UAV_{j}$ be the $j$th UAV, DT of $UAV_{j}$ within time slot $t$ is given as follows
\begin{equation}
\label{deqn_ex1a}
DT^{t}_{UAV_{j}} = \{l^{t}_{UAV_{j}},b^{t}_{j},f^{t}_{j}, c^{t}_{j}, p^{t}_{j,h} \},
\end{equation}
where $l^{t}_{UAV_{j}}$ represents the coordinate $(x^{t}_{j},y^{t}_{j},z^{t}_{j})$ of $UAV_{j}$. $b^{t}_{j}$ is communication bandwidth of $UAV_{j}$. $f^{t}_{j}$ represents the available computing resource of $UAV_{j}$. $c^{t}_{j}$ is the available cache capacity of $UAV_{j}$. $p^{t}_{j,h}$ is the hovering energy consumption of the UAV.
\begin{table}[t]
\center
\caption{List of parameters \label{tab:table1}}
\setlength{\tabcolsep}{1mm}{
\begin{tabular}{l|l}
		\hline
		Parameters & Explanation \\
		\hline	
		$V$ & Set of $UAV$ \\
		\hline
		$M$ & Set of $MED$ \\
		\hline	
		$DT^{t}_{UAV_{j}}$ &  $DT$ of $UAV_{j}$ within time slot $t$ \\
		\hline
		$DT^{t}_{MED_{i}}$ & $DT$ of $MED_{i}$ within time slot $t$ \\
		\hline
		$s^{t}_{i}$ &  Size of tasks \\
		\hline	
		$\vartheta$ &  Computational complexity\\
		\hline
		$p^{t}_{i,tr}$  &  Transmission power\\
		\hline
		$p^{t}_{j,cp}$& Computing power\\
		\hline	
		$p^{t}_{j,h}$ & Hovering power\\
		\hline
		$b^{t}_{j}$ & Communication broadband of $UAV_{j}$ \\
		\hline
		$f^{t}_{j}$ & Available computing resources of $UAV_{j}$\\
		\hline
		$c^{t}_{j}$ &  Amount of caching capability available to $UAV_{j}$ \\
		\hline
		$C^{t}_{ij}$ & Communication capacity of M2U\\
		\hline
		$g^{t}_{ij}$ &  Channel gain between $MED_{i}$ and $UAV_{j}$ \\
		\hline
		$|d^{t}_{i}|$ & Distance between $MED_{i}$ and $UAV_{j}$ \\
		\hline
		$E^{t}_{j,tr}$ &  Communication energy consumption of $UAV_{j}$\\
		\hline
		$E^{t}_{j,cp}$ &  Computing energy consumption of $UAV_{j}$\\
		\hline
		$E^{t}_{j,h}$ &  Hovering energy consumption of $UAV_{j}$ \\
		\hline
		$S^{t}_{V}$ & Set of task from $MED_{i}$ \\
		\hline
		$T^{t}_{V}$ & Delay of complete task \\
		\hline
		$E^{t}_{V,cp}$ & The coalition energy consumption of computing\\
		\hline
		$G$ & Coalition set of UAV\\
		\hline
		$g_{k}$ & $k$th UAV coalition\\
		\hline
		$U^{t}_{gk}$ &  Utility of a coalition \\
		\hline
		$u^{t}_{j}$ & Utility of an UAV \\
		\hline
		$\mu$ & The utilization of computing resource \\
		\hline
	\end{tabular}}
\end{table}

\subsection{Communication Model}
To illustrate the communication capacity between a MED and an UAV coalition (i.e., One-to-many assigning, see Section \ref{sec:4}-A), we assume that one $MED_{i}$ interacts with an UAV coalition in each time slot $t$. The coordinated multiple points transmission technology based on orthogonal frequency division multiple access is introduced in the UAV coalition. And allows UAV to reuse spectrum resources with a total bandwidth of $B^{t}_{max}$ to reduce communication interference. At the same time, other UAVs and MED outside the UAV coalition are in energy-saving mode. Further, the communication capacity between $MED_{i}$ and $UAV_{j}$ in the time slot $t$ is given as follows
\begin{equation}
C^{t}_{ij} = b^{t}_{j} \log \left (1+\dfrac{|g^{t}_{ij}|p^{t}_{i,tr}}{\sigma^{2}} \right),
\end{equation}
where $\sigma^{2}$ is power of Gaussian noise. $b^{t}_{j}$ is the bandwidth that $UAV_{j}$ contributes to the execution of tasks to serve $MED_{i}$, $b^{t}_{j}\in B^{t}_{V}=\{b^{t}_{1},b^{t}_{2},...,b^{t}_{j},...,b^{t}_{n}\}$.
$g_{ij}^{t}$ is the channel gain between $MED_{i}$ and $UAV_{j}$, and its value depends on the distance between $MED_i$ and $UAV_j$, $|d_{ij}|=|(x^{t}_{j}-x^{t}_{i})^{2}+(y^{t}_{j}-y^{t}_{i})^{2}+(z^{t}_{j}-z^{t}_{i})^{2}|^{1/2}$ i.e., $|g_{ij}^{t}|^{2}=|d_{ij}|^{-\gamma}$, path loss $\gamma=4$, then $|g_{ij}^{t}|^{2}=|(x^{t}_{j}-x^{t}_{i})^{2}+(y^{t}_{j}-y^{t}_{i})^{2}+(z^{t}_{j}-z^{t}_{i})^{2}|^{-2}$.

\subsection{Energy Consumption Model}
The energy consumption of $UAV_{j}$ in time slot $t$ mainly includes communication energy consumption, computing energy consumption, pushing energy consumption and hovering energy consumption. $E^{t}_{j,tr}$ is the communication energy consumption of $MED_{i}$ assigning tasks to $UAV_{j}$, and $E^{t}_{j,cp}$ is the computing energy consumed by $UAV_{j}$ to calculate the assigned task. In time slot $t$, the task set $S^{t}_{V}=\{s^{t}_{1},s^{t}_{2},...,s^{t}_{j},...,s^{t}_{n}\}$ is assigned from $MED_{i}$ to the UAV coalition, where $s^{t}_{j}$ is the size of tasks assigned from $MED_{i}$ to $UAV_{j}$, that is, $s^{t}_{j}=s^{t}_{i}$, and its transmission latency $T^{t}_{j,tr}=s^{t}_{j}/C^{t}_{ij}$. At this time, the computing resources contributed by the UAV coalition to complete task of $MED_{i}$ is $F^{t}_{V}=\{f^{t}_{1},f^{t}_{2},...,f^{t}_{j},...,f^{t}_{n}\}$. $f_j^t$ is the computing resource allocated by $UAV_{j}$ to complete task $s_j^t$, then the computing duration is $T^{t}_{j,cp}=\vartheta_{i}^{t}s^{t}_{j}/f^{t}_{j}$, and $f^{t}_{j}>0$. Therefore, the duration for $UAV_{j}$ to complete the task $s^{t}_{j}$ is $T^{t}_{j}=T^{t}_{j,tr}+T^{t}_{j,cp}$, and the delay for the UAV coalition to complete the task is $T^{t}_{V}=max\{T^{t}_{j}|j\in V\}$.

Furthermore, the communication energy consumption spent by $MED_{i}$ to assign the task $s_j^t$ to $UAV_{j}$ in time slot $t$ is $E^{t}_{j,tr}=p^{t}_{i,tr}T^{t}_{j,tr}$. Then the total communication energy consumption of task $S_V^t$ from $MED_{i}$ to the UAV coalition is expressed as
\begin{equation}
E^{t}_{V,tr}=\sum_{j\in V}E^{t}_{j,tr}.
\end{equation}

The computing energy consumption of UAV is $E^{t}_{j,cp}=p^{t}_{j,cp}T^{t}_{j,cp}$. $p^{t}_{j,cp}$ represents the computing energy consumption of $UAV_{j}$ per unit time, taking $p^{t}_{j,cp}=\varepsilon_{j}(f^{t}_{j})^{3}$, where $\varepsilon_{j}$ depends on the parameter of the computing chip, and the dimension is Watt/(cycle/s)$^{3}$\cite{wu2021incentivizing}. Then the computing energy consumption of the UAV coalition is expressed as
\begin{equation}
E^{t}_{V,cp}=\sum_{j\in V}E^{t}_{j,cp}.
\end{equation}

The pushing energy consumption and hovering energy consumption of an UAV are related to factors such as flight speed, weight, wing area, air density, motor speed, and propeller size \cite{liu2023energy, sohail2019energy}. To further simplify the modeling process, it is assumed that the software and hardware of all UAVs are the same. Moreover, the UAV does not change its spatial position during the process of accepting task and executing computation, that is, the UAV's pushing energy consumption is not considered. The hovering power $p^{t}_{j,h}$ per unit time is a constant. The hovering duration depends on the maximum transmission latency and computing duration of the UAV executing tasks, i.e., $T^{t}_{V}$. Then, the hovering energy consumption of $UAV_{j}$ is $E^{t}_{j,h}=p^{t}_{j,h}T^{t}_{V}$, and the total hovering energy consumption of the UAV coalition is expressed as
\begin{equation}
\label{deqn_exla}
E^{t}_{V,h}=\sum_{j\in V}E^{t}_{j,h}.
\end{equation}

Conclusively, $E^{t}_{V}$ represents the corresponding total energy consumption, which is denoted as
\begin{equation}
E^{t}_{V}=E^{t}_{V,tr}+E^{t}_{V,cp}+E^{t}_{j,h}.
\label{E1}
\end{equation}

\section{The Question Statement of Task Assignment}
\label{sec:4}
In this section, we first introduce the task assignment problem in the aerial MEC network and discuss its differences compared with conventional task offloading in MEC networks. Then, the resource coalition cooperation algorithm is proposed.

\subsection{Definition of Task Assignment}
The cooperation of resources from various devices is a fundamental problem in the MEC network. The basic idea of task assignment is to offload the tasks to the appropriate computing devices, which is to improve the utilization of resources and the efficiency of task execution. However, in order to find the optimized task assignment strategy, we are mapping the multiple UAVs to a resource pool. Compared with the existing task offloading in MEC systems, the task assignment in this paper has its own unique characteristics in the aerial MEC network.

\textbf{Peer-to-peer pooling:} All UAVs equipped with various resources in the aerial MEC network are considered equal, contributing to the establishment of a resource pool in the airship. Furthermore, the resources of the UAVs can be used through cooperative games, and their resource capabilities will be uniformly scheduled.

\textbf{Low-to-high scheduling:} The nodes in the aerial MEC network are composed of end users (i.e., MEDs), edge servers (i.e., UAVs), and cloud servers (i.e., airship). The tasks in our proposed network can be executed in low-to-high manner according to task request: 1) one-to-one executing; 2) one-to-many assigning. In this paper, we focus on scheduling resources on multiple UAVs to complete the tasks.

\textbf{One-to-many assigning:} In existing task offloading strategies, the tasks are usually offloaded to a powerful edge server or a remote cloud server for increasing computing efficiency (e.g., One-to-one task offloading, see Fig. \ref{fig01}). However, in the proposed aerial MEC network, the tasks can be assigned to many edge servers.

\subsection{TU Coalition Definition}
In order to formulate the resource cooperation strategy required for task assignment adapted to MED, we model the system based on coalition game theory. The problem of MED task assignment and UAV resource cooperation is described as a TU coalition game model. Furthermore, the TU coalition game in this paper is modeled as a coalition utility optimization problem under the constraints of multiple resource cooperations and task assignment delays. Assume that the TU coalition is represented by a triplet $(V, G, U)$, and $V$ is the set of coalition participants, that is, all UAVs. $G$ is a set composed of any two disjoint UAV coalitions. $U$ is a utility function that uses real functions or real numbers to allocate revenues to coalition participants, that is, coalition utility. Assume that the coalition set $G=\{g_{1}, g_{2},\ldots, g_{k},\ldots, g_{K}\}$ is the result of dividing the set $V$ into multiple coalitions, $g_{k}$ represents the $k$th UAV coalition, and for $\forall k\neq k^{'}$ and $k$, $k^{'}\in \{1,2,\ldots,K\}$, there are $g_{k}$, $g_{k^{'}}\subseteq G$ and $g_{k}\cap g_{k^{'}}=\phi$.

\subsection{TU Coalition Utility}
In this paper, the multi-UAV resource cooperation method based on TU coalition game adopts logarithmic satisfaction function $\log(1+\sum_{j\in g_{k}}s^{t}_{j})$ as the revenue of coalition $g_{k}$ to provide services for $MED_{i}$. Obviously, the utility depends on the total number of tasks in $g_{k}$. At the same time, in order to measure the communication energy consumption caused by UAV cooperative bandwidth resources in $g_{k}$, the communication energy consumption of task assignment is used as the cost for building a coalition. The utility function $U^{t}_{gk}$ of a certain coalition $g_{k}\subseteq G$ in time slot $t$ is as follows
\begin{equation}
U^{t}_{g_{k}}=\varphi\log(1+\sum_{j\in g_{k}}s^{t}_{j})-\varepsilon E^{t}_{g_{k},tr},
\end{equation}
where $\varphi$ is the satisfaction factor of MED with the task assignment services provided by the UAV coalition. $\varepsilon$ denotes the weight of the energy consumption compensation. The utility of the above TU coalition game also applies when any UAV works alone or when there is only one participant in a coalition, i.e., $\lvert g_{k} \rvert=1$.

Next, for any UAV coalition, the information exchange between coalitions is not considered. We consider jointly optimizing the configuration of bandwidth, cache and computing resources of UAVs in the coalition in each time slot $t$, and maximizing the utility of the coalition. Furthermore, the paper comprehensively considers multiple constraints such as UAV's available bandwidth, cache, task size and latency requirements, and then proposes the coalition utility optimization function $F1$ as follows
\begin{equation}
\begin{array}{ll}
F1:&\underset {s^{t}_{j},b^{t}_{j}} {\text{max}} ~~~ U^{t}_{g_k} \\
\text { s.t.: } & C_{1}: 0 \leq b^{t}_{j}\leq b^{t}_{j,max},\\
& C_{2}: \Sigma_{j\in g_{k}} b^{t}_{j} \leq B^{t}_{max}, \\
 & C_{3}: 0 \leq s^{t}_{j}\leq c^{t}_{j,max}, \\
 & C_{4}: \Sigma_{j\in g_{k}} s^{t}_{j} \leq s^{t}_{i},\\
 & C_{5}: T^{t}_{gk}\leq \tau^{t}_{i},
\end{array}
\end{equation}
where $b^{t}_{j,max}$ is maximum available bandwidth of $UAV_{j}$. $B_{max}^t$ is the maximum available bandwidth of the coalition where $UAV_j$ is located. $c_{j,max}^t$ is the largest available cache of $UAV_j$. $C_1$ is the available bandwidth constraint of each UAV in the coalition $g_k$. $C_2$ means that the bandwidth resources available to all UAVs in the coalition cannot exceed the total bandwidth resources of the coalition in which they are located. $C_3$ means that the number of tasks received by each UAV cannot exceed its own maximum available cache. $C_4$ indicates that the total number of tasks in UAV coalition $g_k$ cannot exceed the task assignment amount of $MED_i$ in the current time slot $t$. $C_5$ indicates that the time for coalition $g_k$ to complete the task must meet the latency $\tau_i^t$ requirements of $MED_i$.

\subsection{Coalition Participant Utility}
The process of obtaining the maximum coalition utility is also a process in which each participant in the coalition (specifically the UAV) obtains the corresponding utility by contributing its own available resources. Moreover, the resource contribution of participants is costed by corresponding energy consumption, for which appropriate compensation is required. This paper is designed to use a weighted average to characterize the benefits of coalition participants, and $E_{j, c p}^t$ and $E_{j, h}^t$ serve as the costs incurred by coalition participants in task execution. Furthermore, the utility function of $UAV_j$ within time slot $t$ is as follows

\begin{equation}\label{equ:9}
u_j^t= U^{t}_{g_k} \frac{s_j^t}{\sum_{j \in g_k} s_j^t}-\alpha E_{j, c p}^t-\beta E_{j, h}^t,
\end{equation}
where $\alpha$ and $\beta$ denote the weight coefficients of $UAV_{j}$ computing energy consumption and hovering energy consumption, respectively.

Furthermore, considering the constraints of the computing resources and total task execution duration of $UAV_{j}$ in the time slot $t$, the utility optimization function $F2$ of each participant in the TU coalition is designed as follows

\begin{equation}
\begin{array}{ll}
F2: & \underset {f^{t}_{j}} {\text{max}} ~~~ u^{t}_{j}  \\
\text { s.t.: } & \mathrm{C}_6: 0 \leq f_j^t \leq f^{t}_{j, max }, \\
& \mathrm{C}_7: T_{j, c p}^t \leq \tau_i^t-T_{j, t r}^t,
\label{E01}
\end{array}
\end{equation}
where $f^{t}_{j, max}$ is maximum available computing resources of $UAV_{j}$. $C_{6}$ indicates that the $UAV_{j}$ can decide whether to accept the task assignment, and if it accepts, the UAV cannot be overclocked when executing its task. $C_{7}$ denotes that the time it takes for the coalition to complete the accepted task must meet the latency constraints of the $MED_i$.

In our proposed method, the UAV acts as an aerial base station for providing computing services. Therefore, the utilization of the computing resources of the UAV must be considered when obtaining the maximum participant utility. The utilization of the computing resource $\mu$ by each participant in the coalition is designed as follows

\begin{equation}
\mu = \frac{\tilde{f}_j^t}{f^{t}_{j, max }},
\label{E11}
\end{equation}
where $\tilde{f}_j^t$ is the actual value of the computing resource of the UAV.

\subsection{Problem Conversion and Merging}
To simplify the complexity of solving the coalition utility, this paper transforms the multi-objective nonlinear multi-constraint optimization problem represented by optimization problems $F1$ and $F2$ into a single-objective linear constraint optimization problem. Specifically, considering that $s_j^t$ has been determined in the coalition utility $F1$, that is, the benefit of each coalition participant's utility in the optimization problem $F2$ is constant. At the same time, $E_{i, c p}^t$ will decrease as $f_i^t$ decreases, causing the value of the objective function of the optimization problem $F2$ to increase. Furthermore, under the condition that constraint $C_7$ is satisfied, UAV will reduce the computing resources used to complete the task, and the combined constraints $C_6$ and $C_7$ are as follows
\begin{equation}
\label{11}
\vartheta^{t}_{i} s_j^t /\left(\tau_i^t-T_{j, t r}^t\right) \leq f_j^t \leq f_{j, \max }^t.
\end{equation}

For any coalition participant, if $\vartheta^{t}_{j} s_j^t /\left(\tau_{i, \max }^t-T_{j, t r}^t\right) \leq f_{j, \max }^t$, the computing resources contributed by that participant are $f_j^{t^*}=\vartheta_j^t s_j^t /\left(\tau_{i, \max }^t-T_{j, t r}^t\right)$. If $\vartheta^{t}_{j} s_j^t /\left(\tau_{i, \max }^t-T_{j, t r}^t\right)>f_{j, \max }^t$, then $f_j^{t^*}=f_{j, \max }^t$. Therefore, equation (\ref{11}) can be further simplified as follows

\begin{equation}
\mathrm{C}_8: s_j^t \leq \tau_{i, \max }^t /\left(\frac{1}{C_{i j}^t}+\frac{\vartheta_j^t}{f_{j, \text { max }}^t}\right),
\end{equation}
where $b^{t}_{j}$ in $C^{t}_{j}$ takes the maximum value $b^{t}_{j,max}$. Then, $C_5$ in $F1$ can be further expressed linearly as follows
\begin{equation}
\mathrm{C}_9: s_j^t \leq \tau_{i, \max }^t C_{i j}^t.
\end{equation}

After simplifying multiple constraints and merging functions $F1$ and $F2$, we further obtain the TU coalition game utility as follows
\begin{equation}
\begin{array}{ll}
F3: & \underset {s_j^t, b_i^t} {\text{max}} ~~~ U^{t}_{g_k} \\
\text { s.t.: } &\mathrm{C}_1, \mathrm{C}_2, \mathrm{C}_3, \mathrm{C}_4, \mathrm{C}_8, \mathrm{C}_9.
\label{E02}
\end{array}
\end{equation}

In summary, as the number of coalition participants increases, the cost of interaction between members in a coalition formed by multiple participants will also increase, resulting in a decrease in the utility that participants gain from the coalition. Therefore, how to form an optimal coalition when the coalition utility or participant utility is reduced, this paper explores the TU coalition game method based on DT.
\begin{figure}[!t]
	\centering
	\includegraphics[width=3.5in]{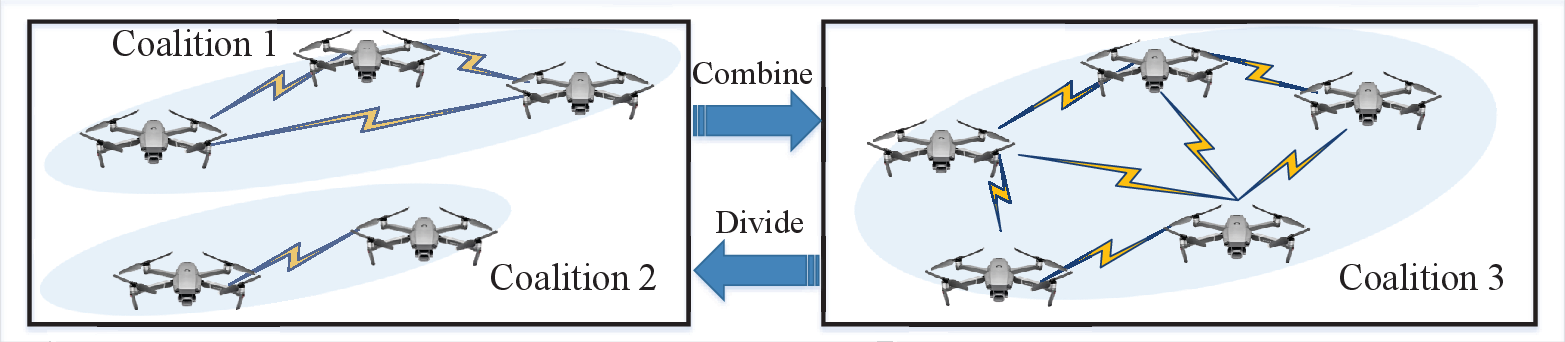}%
	\caption{Combining and dividing rule.}
	\label{fig03}
\end{figure}
\section{Resource Coalition Cooperation Approach based on DT}
\label{sec:5}
In this section, we first introduce a definition of coalition formation, including Pareto optimal, coalition combining, and coalition dividing. Next, we formulate the resource coalition cooperation algorithm based on DT.

\subsection{Coalition Formation Rules}
As coalition participants increase, the number of coalitions formed also increases. However, the existing coalition formation is a centralized solution and requires traversing all coalition participant sets, which is a NP-complete problem \cite{papadimitriou1997np}. To this end, this paper proposes coalition formation rules. The execution process of the rules is shown in Fig. \ref {fig03}. The Pareto optimal rule calculates the utility of participants in different coalitions, realizes distributed dynamic coalition structure, and further sorts coalitions according to their utility. We assume that all participants have their own temporary coalitions, and the proposed Pareto optimal, coalition combining, and coalition dividing rule are as follows

\textbf{Definition 1: (Pareto optimal rule)} Assuming there is a set of coalition participants $v \subset V$, and there are two coalitions $G_1=\left\{g_1^1, g_2^1, \ldots g_{k_1}^1, \ldots g_{K_1}^1\right\}$ and  $G_2=\left\{g_1^2, g_2^2, \ldots g_{k_2}^2, \ldots g_{K_2}^2\right\}$ in $v$. If the utility of $UAV_{j}$ in the coalition set $G_1$ is better than $G_2$, that is, $u_i\left(G_1\right) \geq u_i\left(G_2\right), \forall i \in v$, then the Pareto optimal rule is expressed as $G_1 \triangleright G_2$, if and only if there is at least one $UAV_{j}$ takes a strict inequality.

\textbf{Definition 2: (Combining rule)} Assuming the coalition set $G_3=\left\{g_1, g_2, \ldots, g_{k_b}\right\}$, if there is $G_4=\left\{U_{k_b=1}^{K_b} g_{k_b}\right\} \triangleright G_3$, $G_3$ will be formed into $G_4$, that is, $G_3 \succ G_4$.

\textbf{Definition 3: (Dividing rule)} Assuming the coalition set $G_5=\bigcup_{k_s=1}^{K_s} g_{k_s}$, if there is $G_6=\left\{g_1, g_2, \ldots, g_{k_s}\right\} \triangleright G_5$, $G_5$ will be split into $G_6$, that is, $G_5 \prec G_6$.

According to the above rules, the coalition always maintains the maximum utility in the current time slot and converges to the optimal coalition structure under the current time slot.

\subsection{Resource Coalition Cooperation Algorithm }
\begin{figure}[!t]
	\centering
	\includegraphics[width=3.5in]{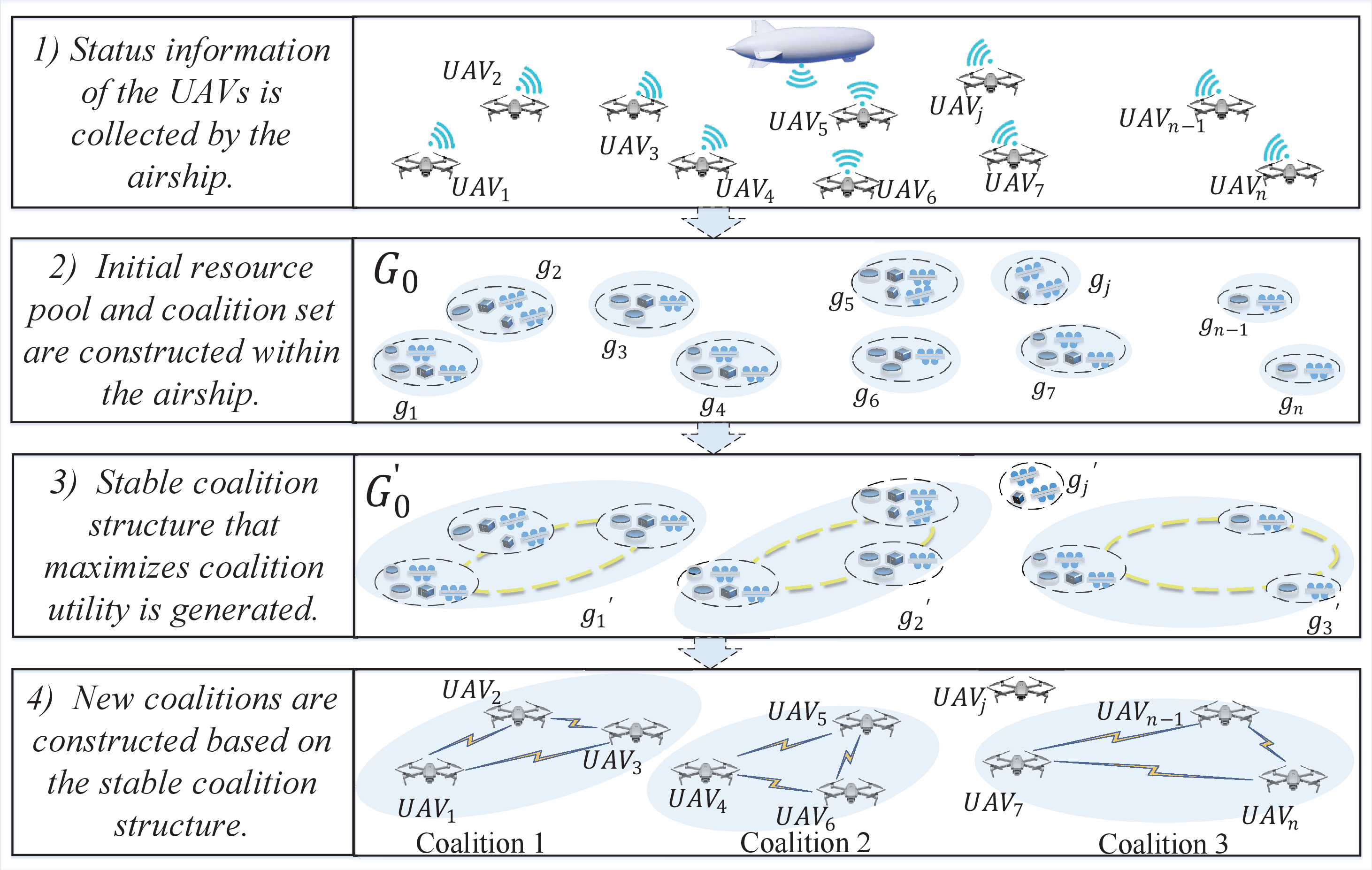}%
	\caption{DT-empowered resource coalition cooperation.}
	\label{fig04}
\end{figure}

Data collection is a key input for building a DT model. We designed the airship to maintain periodic interaction with the UAVs. To obtain the spatial coordinates of UAV and MED, along with task assignment requests, available resources of UAVs, and other relevant data, these elements are used to build the DT model of UAV and MED in the airship, and generate network status $\mathcal N$. Assume that the network status of synchronization time slot $t$ is $n(t)\in N$. Before the next new time slot starts, the current network status $n(t)$ will be transferred to the decision-making module in the airship as the data input of the coalition game based on DT. To simplify the analysis, we define the network status expression as follows
\begin{equation}
n(t)=\left\{\boldsymbol{B}_V^t, \boldsymbol{C}_i^t, D T_{M E D_i}^t, \boldsymbol{F}_V^t, \mathcal{C}_V^t, E n v^t\right\},
\end{equation}
where $\boldsymbol{B}_V^t=\left[b_1^t, b_2^t, \ldots, b_j^t, \ldots, b_n^t\right]$ represents the available bandwidth resource vector in the coalition. $\boldsymbol{C}_i^t=\left[C_{i 1}^t, C_{i 2}^t, \ldots, C_{i j}^t, \ldots, C_{i n}^t\right]$ represents the communication capacity vector between $MED_{i}$ and all participants in the coalition. $\boldsymbol{F}_N^t=\left[f_1^t, f_2^t, \ldots, f_j^t, \ldots, f_n^t\right]$ represents the computing resources available among the participants. $\mathcal{C}_N^t=\left[c_1^t, c_2^t, \ldots, c_j^t, \ldots, c_n^t\right]$ represents the cache capacity available among the participants. $E n v^t=B_{\max }^t$ represents the bandwidth resources currently available to the participants.
\begin{algorithm}[H]
	\caption{Resource coalition cooperation algorithm}\label{alg:alg1}
	\begin{algorithmic}
		\STATE 1: {\bf Input}: Initial DT parameters for MEC and UAVs in airship, i.e., $l^{t}_{MED_{i}}$, $S^{t}_{i}$, $\vartheta^{t}_{i}$, $p^{t}_{i,tr}$, $\tau^{t}_{i}$, $l^{t}_{UAV_{j}}$, $b^{t}_{j}$, $f^{t}_{j}$, $c^{t}_{j}$, $p^{t}_{j,h}$.
		\STATE 2: {\bf Output}: Optimization objective function, i.e., $U^{t}_{g_k}$; The optimization value of resources cooperation, i.e., $s_i^t$, $ b_i^t$, $f^{t}_{j}$.
		\STATE 3: {\bf begin}
        \STATE 4: {\bf for}
		\STATE 5: Execute the \emph{Step 1}, and read the initial DT parameters.
        \STATE 6: Execute the \emph{Step 2}, and initialize the flag of coalition $M\_stab=false$, $S\_stab=false$, $C\_stab=false$.
		\STATE 7: {\bf while} $\sim$$C\_stab$
		\STATE 8: Execute the \emph{Step 3}, and set $K_{b}$ and $K_{s}$.
		\STATE 9: Calculate the optimal utility according to Eq. \eqref{E01} and \eqref{E02}, respectively, and then $M\_stab=true$.
		\STATE 10: The coalition is reconstructed according to the coalition combing and dividing rule, and then $S\_stab=true$.
		\STATE 11: {\bf if} $M\_stab$ \&\& $S\_stab$
		\STATE 12: ~~~~~ $C\_stab=true$
		\STATE 13:        {\bf else}
        \STATE 15:          return to the \emph{Step 3}.
        \STATE 16: {\bf end}
		\STATE 17: {\bf end}
        \STATE 18: {\bf endfor}
		\STATE 19: According to the \emph{Step 4}, coordinate the implementation of the resource coalition cooperation. The user then assigns the tasks to multiple UAVs for computing.
		\STATE 20: {\bf end}
	\end{algorithmic}
\end{algorithm}
The network status $n(t)$ at time slot $t$ is input into the resource scheduler and is further mapped into parameters and constraints in the utility function $F3$. Then, with the help of DT, the virtual and physical mapping of MED and UAV is completed, and combined with the coalition formation rules, a DT-empowered resource coalition cooperation approach is implemented, as shown in Algorithm \ref{alg:alg1}. The execution process of the algorithm is shown in Fig. \ref {fig04}. The detailed steps of this algorithm are as follows

\textbf{Step 1 (Line 1-line 5). Collected status information of the UAVs:} The status information of the multi-UAV is collected by the airship. The DT of the multi-UAV constructs the initial network status $n(t=0)$ based on the collected status information.

\textbf{Step 2 (Line 6). Initial resource pool and coalition set:} The resource scheduler forms each $DT^{t}_{UAV_{j}}$ into a coalition $g_{j}$, so that generates the initial coalition set $G_{0}=\{g_{1}, g_{2},\ldots,g_{j},\ldots, g_{n}\}$.

\textbf{Step 3 (Line 7-line 15). Stable coalition structure with maximization of coalition utility:} Based on the status information of the task assignment, all coalitions maximize the utility of each coalition participant by combining or dividing. When several coalitions satisfy the rules of combining, these coalitions will be reorganized into a larger coalition, i.e., $G^{'}_{0}=\{g^{'}_{1}, g^{'}_{2}, g^{'}_{3}, g^{'}_{j}\}$. When a coalition satisfies the dividing rule, the coalition is split into smaller coalitions.

\textbf{Step 4 (Line 19). Constructed new coalitions:} Resource scheduler selects the coalition with the optimal utility as the input for resource coalition cooperation, and then determines the participants in the coalition. When the current time slot ends, immediately abandon the original coalition construction and return to the \emph{Step 2}.
\section{Performance Analysis}
\label{sec:6}
In this section, the stability and convergence of coalition are illustrated. Then, we define the fidelity of DT model and analyse its deviation.

\subsection{Stability Definition and Proof}
\textbf{Definition 4: (Stability)} To prove that the resource coalition cooperation algorithm can still converge to a stable status after a limited number of coalitions combine and divide. Therefore, we use the ${\mathbb{D}}_{hp}$ stability coalition method \cite{chen2020joint} to analyze the stability of the algorithm proposed in this paper. If there is a stable coalition set $G=\{g_{1}, g_{2}, \ldots, g_{k}, \ldots, g_{K}\}$, $g_k$ is arbitrarily divided into multiple sub-coalitions $\left\{g_1, g_2, \ldots, g_q, \ldots, g_Q\right\}$, and then $\left\{g_1, g_2, \ldots, g_q, \ldots, g_Q\right\} \ntriangleright g_k$. And if $\forall q \in\{1,2, \ldots, q, \ldots, Q\}$, $Q \leq K$, then there is $\cup_{q \in Q} g_q \ntriangleright \left\{g_q \mid q \in Q\right\}$, where $\ntriangleright$ means that the object on the left side of the symbol is not better than the object on the other side of it. If a coalition set meets the above two conditions at the same time, it means that the coalition is ${\mathbb{D}}_{hp}$ stable.

\textbf{Proof:} The system model in this paper is a discrete periodic synchronization time slot system, and network status updates only occur at the end of the synchronization time slot. We want to prove that the TU coalition game algorithm proposed in this paper can converge to a unique and ${\mathbb{D}}_{hp}$ stability coalition within a limited number of iterations. The specific proof is as follows

\emph{a)} If $\forall g_{k}\in G$ is split into $\{g_{1}^{'}, g_{2}^{'}, \ldots, g_{q}^{'}, \ldots, g_{Q}^{'}\}$ and there is $\{g_{1}, g_{2}, \ldots, g_{q}, \ldots, g_{Q} \}\rhd g_{k}$. Then according to the dividing rule coalition $G \prec G^{'}=\{g_{1}, g_{2}, \ldots, g_{1}^{'}, g_{2}^{'}, \ldots, g_{q}^{'},\ldots, g_{Q}^{'}, \ldots, g_{K}\}$, the split result is inconsistent with the coalition set $G$.

\emph{b)} If there is $\cup_{q \in Q} g_q \rhd \left\{g_q \mid q \in Q\right\}$, then according to the coalition combining rules, $G \succ G^{''}= \left\{ \cup_{q \in Q} g_q \right\}\cup \left\{g_q \mid q \in \{ 1,2,\ldots, K\} \backslash Q\right\}$ can be obtained. However, the combining result is contradictory to $G$.

The above assumptions \emph{a)} and \emph{b)} further illustrate that as long as the coalition is stable, then the coalition must be the only coalition structure in the current network status.

\subsection{Convergence Analysis}
Suppose the initial coalition is $G_0$, and the set of stable coalitions after the $x$th optimization is $G_x$. Then, the coalition combining and dividing rules are executed, and $G_x$ is formed after the $(x+1)$th combining or dividing of the coalition set $G_{x+1}$ that has not been iteratively traversed. Similarly, the initial coalition $G_0$ becomes $G_X$ after $X$ rounds of combine or divide iterations. Obviously, for a set $V$ composed of $n$ coalition participants, under the limitation of the objective function of optimizing coalition utility with multiple constraints, the number of coalitions constructed must be enumerable and can be solved using Bell numbers. At the same time, with the optimization iteration of the coalition, the TU coalition game method proposed in this paper can also tend to converge in a limited number of iterations, forming a stable coalition in the current time slot, thereby providing task assignment decisions for MED.

\subsection{DT Deviation Analysis}
High-fidelity DT model relies on real-time interactions between physical space and virtual space. The computing resources of the UAV change dynamically, so that its computing resource $f_{j}$ cannot be accurately obtained. Therefore, the actual computing resource of $UAV_{j}$ is ${\tilde{f}}_{j} = f_{j} + \mathrm{\Delta}f_{j}$, where $\mathrm{\Delta}f_{j}$ is the deviations. In order to measure the fidelity of DT model reflecting physical object parameters, we define the fidelity ${{\mathcal{F}}_{\mathcal{N}}}=1-\left| \Delta \mathcal{N}/\mathcal{\tilde{N}} \right|$ as the deviation between the estimated value in the DT model and the true value of the corresponding physical object status quantity \cite{liu2023review}. That is, the smaller the deviation between the DT model parameters and the physical real values, the higher the fidelity of the DT model generated in the formal update phase. ${{\mathcal{F}}_{{{f}_{j}}}}=1-\left|\Delta {{f}_{j}}/{{\tilde{f}}_{j}} \right|$ is the fidelity of computing resource.

To investigate the effect of the deviations between the estimated values of $f_{j}$ and the actual values of ${\tilde{f}}_{j}$, the actual computing duration required for $UAV_{j}$ to perform task assignment is expressed as

\begin{equation}
{\tilde{T}}_{j,cp} = \vartheta_{i}s^{t}_{j}/{{\tilde{f}}_{j}}.
\label{E2}
\end{equation}

The deviation on the total energy consumption can be derived from \eqref{E1} and \eqref{E2} as follows
\begin{equation}
\tilde{E}_{V}^{t}=E^{t}_{V,tr}+\sum_{j\in V}\left(p^{t}_{j,cp}{\tilde{T}}^{t}_{j,cp}+p^{t}_{j,h}{\tilde{T}}_{V}^{t}\right),
\label{E3}
\end{equation}
where ${\tilde{T}}^{t}_{V}=max\{T^{t}_{j,tr}+{\tilde{T}}^{t}_{j,cp}|j\in V\}$.


\section{Simulation Results and Discussion}
\label{sec:7}
This section is to verify the performance advantages of the resource coalition cooperation and task assignment strategy proposed in this paper. We perform numerical simulation and analysis in aspects such as effectiveness of DT and GM, resource utilization, energy consumption, and coalition utility.
\begin{table}[!t]
	\caption{ Simulation experiment parameter setting \label{tab:table2}}
	\centering
	\begin{tabular}{|c|c|}
		\hline
		Parameters & Values \\
		\hline
		Number of $UAV$ & $[5,30]$ \\
		\hline
		Flight area & 1 km$^{2}$  \\
		\hline
		Hovering energy consumption per unit time & 168 J \\
		\hline
	    Total task size & $[5,25]$ Mbyte \\
		\hline	
		Computational complexity & $[50,300]$ cycle/bit \\
		\hline
		Maximum transmission power &$[50,100]$ mW \\
		\hline
	    Maximum bandwidth resources &$[1,5]$ MHz \\
		\hline	
		Maximum cache resources & $[1,2]$ Mbyte \\
		\hline	
		Maximum tolerable latency & $[150,500]$ ms \\
		\hline	
		Computing of energy consumption coefficients & $[1,2.5]$ W/(cycle/s)$^{3}$ \\
		\hline
		Noise power & -110 dBm \\
		\hline
		Maximum computing resources & $[4,10]$ GHz \\
		\hline
		Maximum multiplexing bandwidth & 16 MHz \\
		\hline
	\end{tabular}
\end{table}
\subsection{Simulation Settings}
Aim to make the experiment better focused on the resource coalition cooperation strategy proposed in this paper, we select an area of 1 km $*$ 1 km set that is covered by an airship and multiple UAVs in the data set. The simulation environment is mainly for the needs of densely populated urban high-rise buildings. The existing most urban high-rise buildings are below 800 m. Therefore, we consider all UAVs flying at a fixed altitude of 800 m. Furthermore, in addition to cooperating with the UAVs below, the airship will also carry some observation payloads for air-to-ground monitoring. At this time, its own safety factors must also be considered. For example, in the case of sudden volcanic disasters, the height of the smoke column can reach hundreds or even thousands of meters. In this way, we set an airship to hover at a fixed altitude of 2000 m. The transmission channels of M2U and U2A in this experimental design of an aerial MEC network adopt the Rayleigh fading model. The number of time slots is set to 500 and each slot has a maximum duration $t$ = 60 s. The DTs of UAVs and MEDs are established on the airship. The DTs of MEDs reflect the dynamic demand and supply of resources by the UAVs. Hence, we turn the non-convex resource optimization problem into a convex optimization problem with linear constraints that can be solved using conventional optimization tools such as CVX. Moreover, we feed about two hundred labeled strategies to the diffusion model and train the denoising process of U-Net with maximum-likelihood estimation \cite{huang2023ai}. The tasks were generated randomly on the MEDs and assigned to UAVs with airship assistance. We perform simulation experiments in the platform of MATLAB R2019a. The PC is equipped with the 64-bit Microsoft Windows 10 operating system, 16 GB RAM, and Intel(R) Core(TM) i5-8250U CPU. To evaluate the efficiency of our proposed solution, the utilization of computing resource \eqref{E11}, the total energy consumption Eq. \eqref{E1}, and the participant utility Eq. \eqref{equ:9} of our proposed scheme with these of benchmark schemes are compared under four different settings, i.e., the computational complexity $\vartheta^{t}_{i}$, the number of UAV $n$, and the number of time slots $\tau$. The powers of UAV hovering and other parameters are listed in Table \ref{tab:table2} \cite{sohail2019energy}.

Moreover, in order to reflect the advantages of proposed method, we compare the performance of the proposed approach with the following alternatives, i.e., the cooperation-based grand coalition method \cite{9390406} and the noncooperation-based Nash equilibrium method \cite{chen2020joint} as the benchmark method.

\textbf{Grand coalition:} Grand coalition means all UAVs cooperate with each other to form a multi-UAV system to provide all-to-one task-executing services. In this way, the coalition enhances the quality of services. However, it can also lead to poor utilization of UAV resources. And this approach does not pay attention to the utilization of all resources.

\textbf{Nash equilibrium:} Nash equilibrium means the UAV independently provides task offloading services for MED, requiring a careful evaluation of its utility in multi-UAV. Each UAV provides task offloading services independently, requiring careful evaluation of the inter-UAV relationships. In other ways, additional information interaction among multiple UAVs is required when generating resource cooperation strategies.

\begin{figure}[!t]
	\centering
	\includegraphics[width=2.8in]{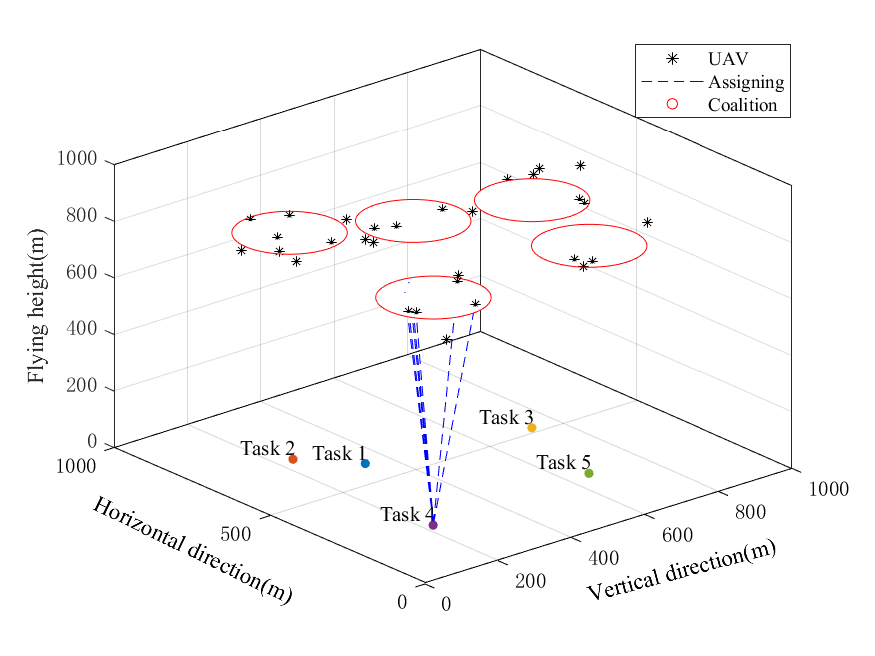}
	\caption{The coalition combining and task assignment.}
    \label{fig_5}
\end{figure}
\begin{figure}[!t]
	\centering
	\includegraphics[width=2.8in]{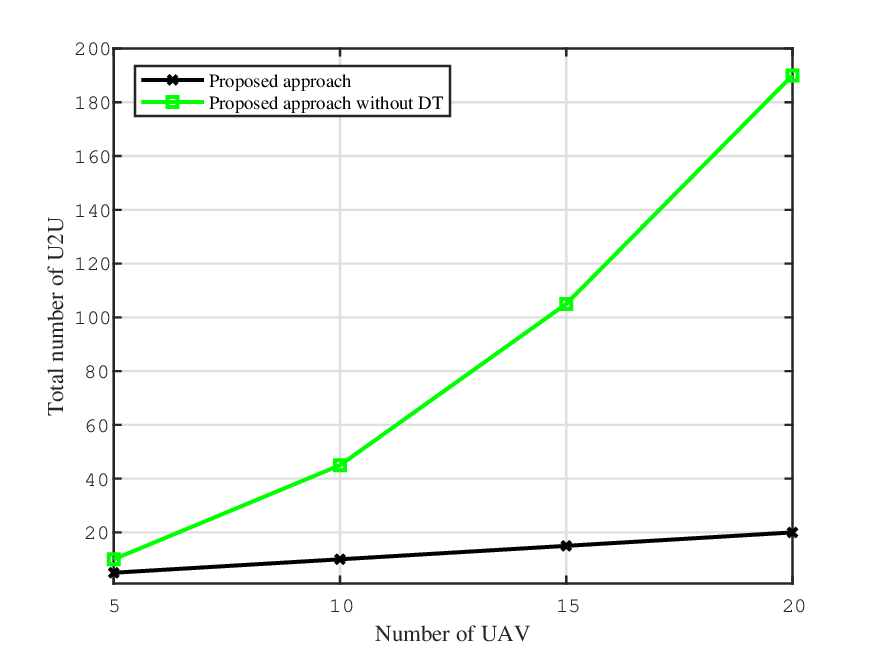}
	\caption{The total number of U2U with different numbers of UAVs.}
	\label{fig_5-3}
\end{figure}
\subsection{Implementation of One-to-Many Task Assigning}
Fig. \ref{fig_5} shows the coalition combining and task assignment results in aerial MEC network. According to the coalition formation rules, the coalition participant set goes through multiple rounds of calculation and select the optimal coalition participant utility. Then, all coalitions are combined and divided multiple times to realize that the coalition structure changed with the task assignment requirements and then autonomously select coalition participants. After applying the above rules, all UAVs achieve stable coalitions. These results show the feasibility of our method.

\subsection{Effectiveness Analysis of DT and GM}

\begin{figure}[!t]
	\centering
	\includegraphics[width=2.8in]{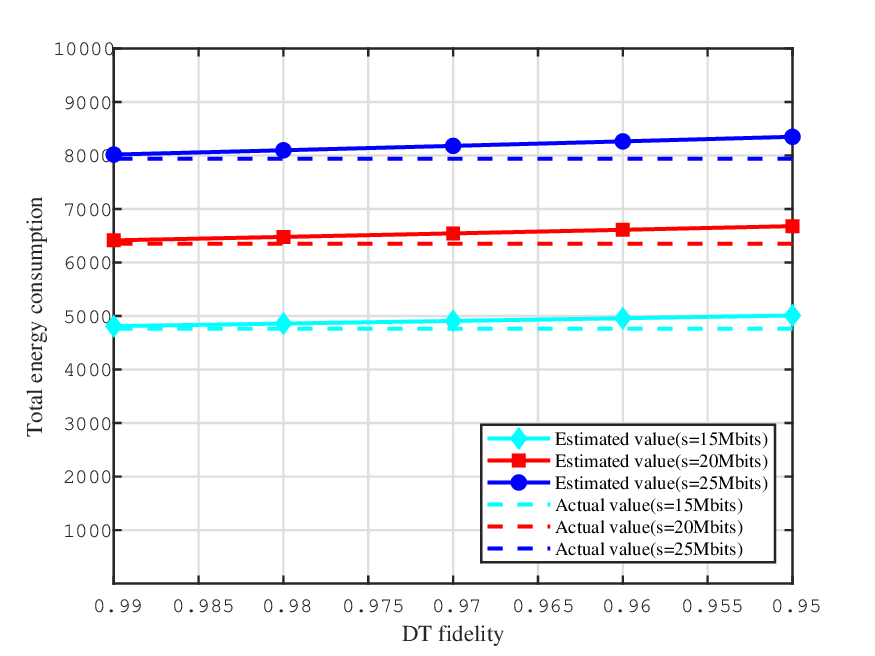}
	\caption{The total energy consumption of the coalition with different DT fidelity.}
	\label{fig_5-4}
\end{figure}
\begin{figure}[!t]
	\centering
	\includegraphics[width=2.8in]{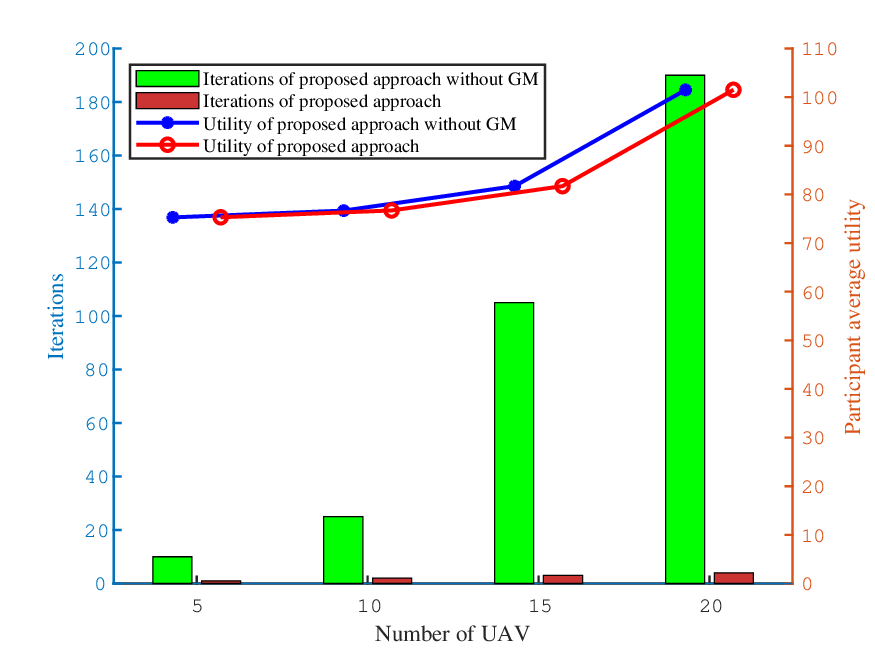}
	\caption{The effectiveness of the proposed approach in terms of iteration.}
	\label{fig_5-5}
\end{figure}
In Fig. \ref{fig_5-3} and Fig. \ref{fig_5-4}, we analyze the impact of the appearance of DT on the aerial MEC network's performance. As can be observed, in Fig. \ref{fig_5-3}, the performance measured by the energy consumption of the system under DT-empowered is significantly better than that without DT. The reason for this impact is that the states of each MED and UAV are stored in DT, and no additional information interaction is required when generating resource cooperation strategies. Then, we measure the total energy consumption of the coalition over the varying DT fidelity of the computing resource in Fig. \ref{fig_5-4}. It can be clearly seen that when the task size is constant, the total energy consumption is correlated with the DT fidelity. This can be explained from Eq. \eqref{E3} that a large positive fidelity means the estimated value of DT is worse than the actual value, and thus the actual energy consumption is less than the estimated value. On the other hand, under the condition that the DT fidelity is constant, as the task size increases, the energy consumption will increase. This is because when computing tasks increase, more data needs to be transmitted from MED to UAVs due to the latency constraints, which in turn consume additional energy for data transmission.

Fig. \ref{fig_5-5} compares the iterations obtained by two types of resource coalition cooperation strategies as the UAV number increases. It shows that the proposed resource coalition cooperation approach needs only less iteration computing and achieves the same utility compared to the approach without GM. The reason is that our proposed approach combines the autonomous learning and decision-making capabilities of the GM. The advantage of this is that a preliminary customized resource cooperation strategy can be generated in advance, which reduces the iterations and improves the timeliness of decision-making.

\subsection{Resource Utilization Results and Analysis}
\begin{figure}[!t]
	\centering
	\includegraphics[width=2.8in]{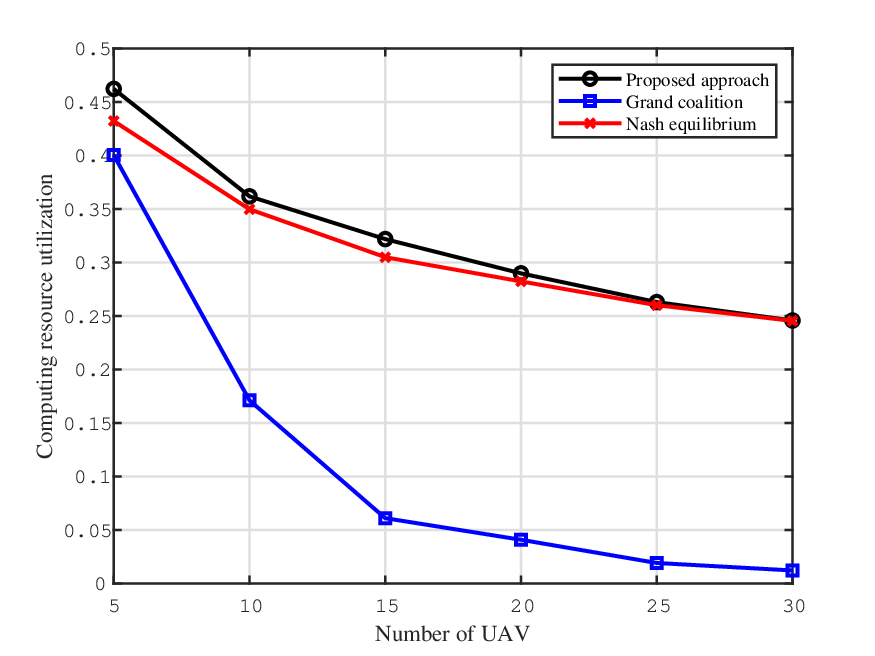}
	\caption{The utilization of computing resources for different numbers of UAVs.}
	\label{fig_7}
\end{figure}
\begin{figure}[!t]
	\centering
	\includegraphics[width=2.8in]{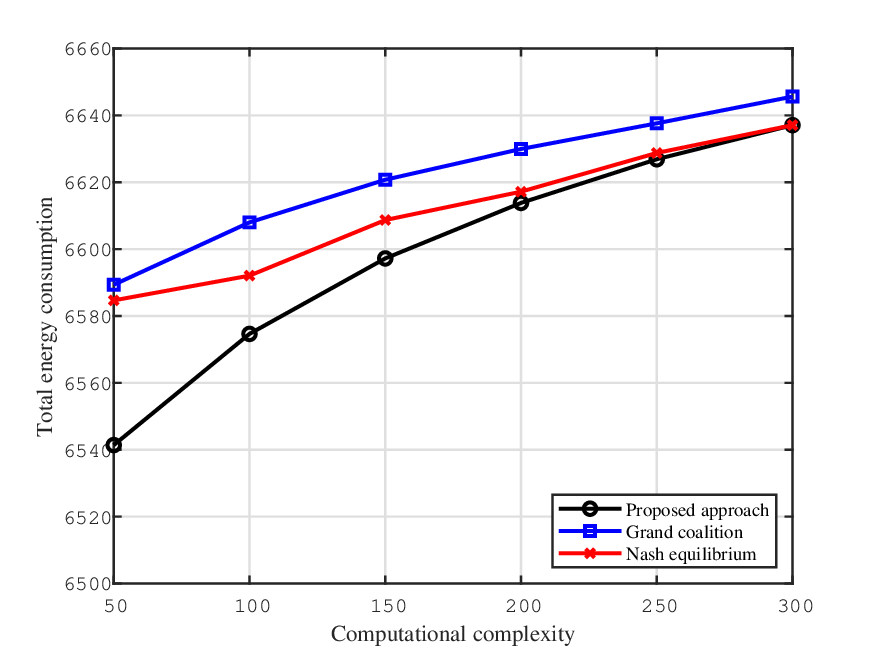}
	\caption{The total energy consumption with different computational complexities.}
	\label{fig_9}
\end{figure}
\begin{figure}[!t]
	\centering
	\includegraphics[width=2.8in]{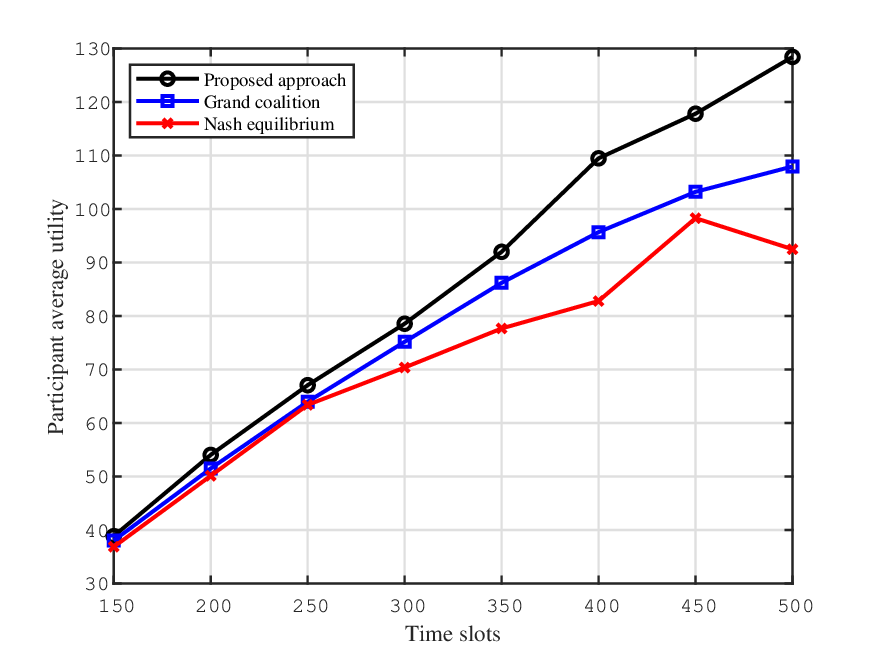}
	\caption{The utility of the coalition participants under different time slot constraints.}
	\label{fig_10}
\end{figure}
The utilization of computing resources for different numbers of UAVs are given in Fig. \ref{fig_7}. It can be seen that the coalition game approach proposed in this paper always outperforms the grand coalition and Nash equilibrium approaches. The reason for this is that the coalition game approach requires only some UAVs to act as coalition participants in providing services for task assignment, while the computing resource cooperation strategies are iteratively optimized. In contrast, the grand coalition approach is to provide all UAVs with services for task offloading, and the resources allocated to each UAV are time-invariant, and the various onboard resources cannot be dynamically adapted to the changes of the task. At the same time, as the number of UAVs participating in the task grows, the resource utilization of the grand coalition approach is inferior to that of the proposed approach. Unlike the previous two cooperative game methods, the Nash equilibrium focuses on maximizing the benefits of each participant's coalition and does not form resource collaboration between them. Therefore, although the multi-UAV resource utilization based on the Nash equilibrium improves compared to the grand coalition, it lacks an effective resource cooperation mechanism between multi-UAVs, and with the incremental increase in the number of UAVs, the resource utilization based on the method does not improve significantly.

\subsection{Energy Consumption Results and Analysis}
The results of the total energy consumption of all UAVs for different computational complexities are given in Fig. \ref{fig_9}. From this figure, it can be observed that the total energy consumption of all UAVs gradually increases as the threshold value of the task execution latency increases. This is because as the computational complexity increases, the UAV is required to invest more arithmetic power in processing and requires more computation latency, which in turn leads to the computing energy consumption being proportional to the computational complexity, and obviously, the total energy consumption is also elevated. At low computational complexity ($\vartheta$ = 50), given that the coalitional gaming approach requires only partial UAV participation to accomplish this type of task, the total energy consumption has a significant advantage over the remaining two types of approaches. However, as the computational complexity of the task increases ($\vartheta$ = 300), more UAVs are required to participate in the task execution, which leads to a decrease in the energy performance gap between these three types of methods. The coalition game and the Nash equilibrium-based methods are based on the premise of improving the utility of the UAVs, and thus the energy performance of the two approaches converges to the same level.

\subsection{Utility Results and Analysis}
Fig. \ref{fig_10} gives the average utility results of the coalition participants under different time slot constraints. From this figure, it can be seen that the average utility of the coalition participants rises as the threshold value of the task execution delay increases. This is because the longer the execution latency of the task, the more resources the coalition participants are willing to put into the task execution, and accordingly, the number of tasks that can be accomplished by each UAV will increase. Then, according to the coalition participant utility Eq. \eqref{equ:9}, the coalition participant occupies a larger proportion of the coalition utility allocation, and each coalition participant's utility will be larger. However, for the grand coalition approach, multiple UAVs are all participants in the task offloading service, which results in fewer tasks for each participant and inevitably leads to less allocated utility. At the same time, the computational and hovering energy consumption of each UAV depends on the maximum transmission latency and computing duration decision of the UAV with the task. Multiple UAVs will be in idle status, leading to useless hovering and lower utility for the participants. While the Nash equilibrium itself is a noncooperative game method, the utility of each UAV participant is not determined by all the participants. It is in a status of mutual constraints, so that the utility of each participant decreases instead of increasing when its utility reaches a certain peak ($\tau$ = 450), which reflects that the noncooperative game-based method is suitable to be used within a certain constraint range.

\section{Conclusion}
\label{sec:8}
In this paper, we propose a DT-empowered task assignment based on resource coalition cooperation with a GM in an aerial MEC network that realizes the mapping of physical space resources to a resource pool in virtual space. Furthermore, to improve the energy consumption of each coalition participant as little as possible and further adapt to the MED task assignment request to maximize the utility of the coalition participants, our proposed method optimizes the bandwidth and computing resource cooperation between UAVs by establishing a coalition game model with a GM so as to realize the sharing and optimal deployment of different resources. The advantages of a coalition game with a GM can be leveraged to improve iterative efficiency in decision-making. The simulation results show that, compared with benchmark algorithms, the proposed method increases resource utilization and the average utility of the coalition participants, which not only realizes the dynamic matching of resource reserves of an aerial MEC network with demand changes but also reduces energy consumption. The proposed method can jointly optimize the utility of UAVs and the interactions of U2U.

In future work, we intend to explore the collaboration between multiple heterogeneous UAVs, while the other is the application of generative models in specific tasks, such as assisting the generation of oblique photography images and wireless spectrum maps.

%
%
%
%

\end{document}